# Photo switching of protein dynamical collectivity

M. Xu, D. K. George, R. Jimenez, and A. G. Markelz

## Abstract

We examine changes in the picosecond structural dynamics with irreversible photobleaching of red fluorescent proteins mCherry, mOrange2 and TagRFP-T. Measurements of the protein dynamical transition using terahertz time-domain spectroscopy show in all cases an increase in the turn-on temperature in the bleached state. The result is surprising given that there is little change in the protein surface, and thus the solvent dynamics held responsible for the transition should not change. A spectral analysis of the measurements guided by quasiharmonic calculations of the protein absorbance reveals that indeed the solvent dynamical turn-on temperature is independent of the thermal stability/photostate however the protein dynamical turn-on temperature shifts to higher temperatures. This is the first demonstration of switching the protein dynamical turn-on temperature with protein functional state. The observed shift in protein dynamical turn-on temperature relative to the solvent indicates an increase in the required mobile waters necessary for the protein picosecond motions: that is these motions are more collective. Melting-point measurements reveal that the photobleached state is more thermally stable and structural analysis of related RFP's shows that there is an increase in internal water channels as well as a more uniform atomic root mean squared displacement. These observations are consistent with previous suggestions that water channels form with extended light excitation providing $O_2$ access to the chromophore and subsequent fluorescence loss. We report that these same channels increase internal coupling enhancing thermal stability and collectivity of the picosecond protein motions. The terahertz spectroscopic characterization of the protein and solvent dynamical onsets can be applied generally to measure changes in collectivity of protein motions.

## I.  INTRODUCTION

Fluorescent proteins provide vital insight into biochemical processes. Since the discovery of green fluorescent protein (GFP) in *Aequorea victoria* researchers have generated a large number of mutants that enable access to a rainbow of excitation and emission wavelengths. The power of these proteins is the formation of their chromophore autocatalytically from three sequential residues (1), thus enabling the incorporation of the fluorescent tag into a targeted protein's expression.  All fluorescent proteins (FPs) share several structural features:  an 11-stranded β-barrel with an internal distorted α-helix to which the chromophore is attached (2, 3). The chromophore removed from the protein has a poor emission yield, four orders of magnitude lower compared to when the chromophore is inside the β-barrel (4).  The β-barrel structure protects the chromophore from external quenchers and inhibits the dark state conversion through a cis to trans isomerization or light-induced protonation/ deprotonation (5, 6).  To achieve higher transmission through tissues, longer wavelength excitation red FPs (RFPs) have been developed. Unfortunately, the application of RFPs to imaging is limited by their higher tendency to be irreversibly photobleached. Though this photobleaching is useful for certain specialized microscopic techniques, such as the fluorescence recovery diffusion measurements, it is more often a bottleneck for imaging applications, especially for those demanding high irradiance illuminations such as single molecule microscopy.  The mechanisms that lead to this decrease in photostability are a current area of study.  Here we examine the changes in the picosecond structural dynamics of several red fluorescent proteins to see if and/or how dynamical changes may be associated with photobleaching (7, 8) .

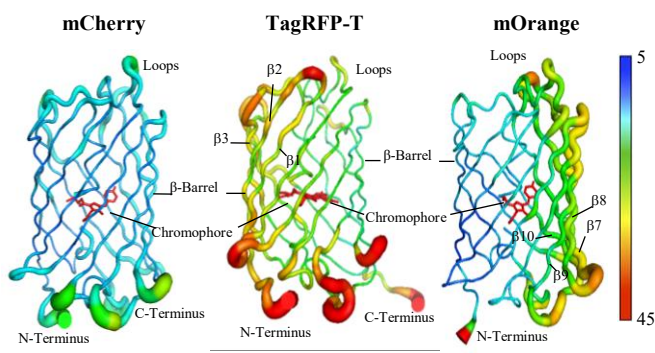

Figure 1.   Structures of mCherry (2H5Q.pdb), TagRFP-T (5JVA.pdb), and mOrange (2H5O.pdb). Note that mOrange is shown as the crystal structure of mOrange2 has not been reported yet.  B-values recorded in each PDB file are quite different for such similar structures due to various techniques, but the width and color of the ribbon in the figure reflect the average B-value dependent on residues to focus on relative B-valve distribution, which has been normalized to each RFP's overall averaged value.

Three engineered monomeric RFPs were studied: mCherry (9), mOrange2 (derivatives of DsRed from *Discosoma* sp.), and TagRFP-T (a derivative of EQFP611 from *Entacmaea quadricolor*) (5). The structures are shown in Figure 1 (10, 11). These three proteins have substantially enhanced resistance to both irreversible and reversible photobleaching under different illumination conditions (12). Techniques to characterize structural flexibility have increasingly focused on the important picosecond time scale which corresponds to librational motions of the solvent and amino acid sidechains as well as the long-range intramolecular vibrations associated with conformational change (13-16). Here we use solution phase temperature dependent terahertz (THz) time-domain spectroscopy, which is a benchtop optical measurement with small sample size requirements that enables a systematic comparison between different mutants as a function of photoexcitation.

Comparing the THz measurements to CD measurements of thermal denaturing, the rate of increase in THz absorbance with temperature is found to be inversely proportional to the thermal stability. Molecular modeling using quasiharmonic vibrational analysis suggests the protein dielectric response can be modeled by a Lorentzian centered at ~ 0.5 THz, (20 cm$^{-1}$, 2.5 meV). A spectral decomposition of the measured frequency dependent absorbance into the protein dielectric response and the bulk water intermolecular stretching mode at 5.3 THz (177 cm$^{-1}$, 22 meV) (17) shows that while the protein dynamical transition is dependent on the specific protein and photobleaching, the water has a temperature dependence that is independent of the specific RFP or its photostate. This is the first report of the protein dynamical turn-on changing independently of the solvent turn-on.

## II. METHODS
### A. Sample preparation

The RFPs were expressed and purified as discussed in Ref. 12. The original concentration of the protein solution was 4 mg/ml in 15 mM MOPS buffer with 100 mM KCl at pH 7.0. It was concentrated to 80 mg/ml in the same dialysis buffer to increase the optical density at THz frequencies, using Eppendorf 5424 Microcentrifuge with Millipore Amicon 10K centrifugal filter.

A Dolan-Jenner Fiber-Lite Illuminator light source and an Ocean Optics High Resolution Spectrometer were utilized to characterize the absorption and emission spectra. The spectral peak shifts as a function of concentration (Figure S6 in Supplementary Information). In addition, it has been verified that each sample completely loses fluorescence after 30 seconds under the 4 W/cm$^2$ 532-nm illumination condition. No protein aggregation occurs in the highly concentrated solution, confirmed by the dynamical light scattering (DLS) measurements using the Zetasizer Nano ZS90 system.

### B. THz measurements

A conventional home-built THz time-domain spectroscopy system was employed to measure the optical absorption of RFPs within the THz range (0.2-2.2 THz) (18, 19). A solution cell with 100-micron thick sample was placed in a gas exchange cryostat and cooled with liquid nitrogen. A silicon diode thermal sensor monitored the temperature directly adjacent to the sample aperture. After the photoactive state was measured, each sample was completely photobleached by 4 W/cm$^2$ 532-nm illumination and at room temperature, then cooled down to 80 K for photobleached state temperature dependent measurements. The photobleaching was confirmed by monitoring the fluorescence while illuminating.

### C. CD measurements

Thermal stability was measured using a Jasco J-715 Spectropolarimeter for temperature dependent far-UV CD spectra. Samples were prepared at 5 μM concentration in 10 mM sodium phosphate buffer at pH 7.5 and sealed in a 10-mm path-length quartz cell. The sample was equilibrated at each temperature for 2 minutes, then 10 scans at 0.1-nm resolution were accumulated and averaged. The scan speed was set to 50 nm/min with 4 sec detector response time. The temperature was gradually increased from 20 °C to 100 °C, in steps of 2 °C, using a Jasco PTC 348 WI temperature controller. The spectrum of a buffer blank was used as the reference. The melting temperatures were determined from the peak of the ellipticity temperature gradient, $d\theta/dT$, measured at 218 nm. In the case of photobleached mCherry, the increase in the melting onset is so large that $d\theta/dT$ maximum is not yet attained by the highest temperature measured (373 K). We estimate the lower bound of the photobleached mCherry melting temperature from the midpoint between the low temperature ellipticity value and the value at the highest temperature measured, 373 K (see Figure S1).

### D. MD calculations

The temperature dependent absorbance spectra were calculated using quasiharmonic mode analysis and dipole autocorrelation. The MD trajectory of the net dipole was calculated with the CHARMM 39 (20), and CHARMM 36 empirical force field (21). The special formation by the auto-oxidation and cyclization of three to four amino acid residues requires the chromophore to be additionally parameterized (22). We applied the residue topology and parameter files for the chromophore used in Ref. 22 to calculations presented here. The initial molecular structures were obtained from the Protein Data Bank, mCherry (PDB: 2H5Q) and TagRFP-T (PDB: 3T6H). The mOrange2 structure was obtained by modifying mOrange (PDB: 2H5O) with Q64H/ F99Y/ E160K/ G196D mutations implemented using the CHARMM-GUI (23).

The CHARMM/TIP3P parameter set was used to model the fully solvated protein system starting from the X-ray determined structure. Each structure was first minimized at zero temperature using steepest descent method followed by the adopted basis Newton-Raphson method until the total energy gradient reached below $10^{-7}$ kcal Å$^{-1}$. The energy-minimized systems were then neutralized by adding potassium or chloride ions, randomly distributed throughout the volume.

MD simulations were performed with an integration time step of 1 fs. Each system was heated from 100 K to 300 K, with a linear gradient of 1 K/ps. The systems were equilibrated for 10 ns in the isobaric-isothermal (constant pressure-temperature, CPT) ensemble. 4-ns long production trajectories were further performed under the CPT condition at 300 K. The dipole moments about the center of geometry were obtained and the absorption intensity was determined from the power spectra of the dipole-dipole autocorrelation function.

## III. RESULTS

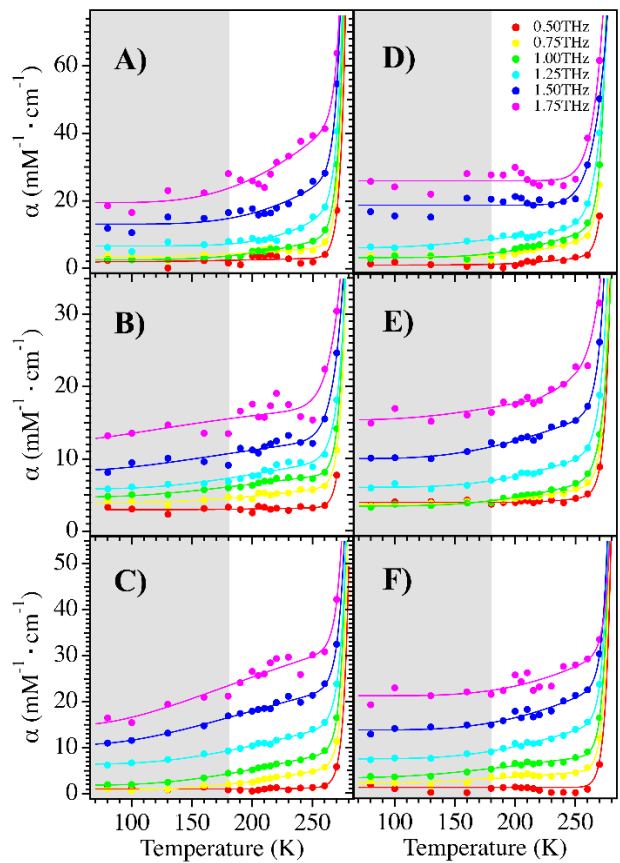

Figure 2. THz molar absorptivity for photoactive (left) and photobleached (right) RFP hydrated samples: TagRFP-T (A, D), mCherry (B, E), and mOrange2 (C, F), with interpolated lines as a guide for the eye, as a function of temperature at different frequencies. Note that vertical axis scale varies with different THz response: TagRFP-T 0-75, mCherry 0-35, mOrange2 0-55 mM$^{-1}$cm$^{-1}$. The low temperature region is indicated by grey shading.

In Figure 2 we show the molar absorption coefficient $\alpha(\omega,T)$ for the three RFPs at several THz frequencies for both the functional and photobleached states. Both protein resonance and water absorption contribute to the spectra while water dominates. Previously it has been stated that the THz absorbance follows root-mean-square-deviation (RMSD) measured by neutron scattering in the same energy range dependent on temperature and hydration. We identify two temperature regimes for the molar absorptivity: a low temperature harmonic regime (80-200 K), and a high temperature anharmonic regime (250-270 K). Below 200 K, $\alpha(\omega,T)$ increases linearly with temperature indicating that the protein vibrations in the THz range are harmonic, which is consistent with the $q^2$ dependent boson peak and linearly increasing root-mean-square-deviation (RMSD) shown in inelastic neutron scattering (INS) (24). At ~200 K the absorbance rapidly increases as the dynamics move to the anharmonic regime. The low temperature slopes differ for the three proteins, with TagRFP-T having the largest slope and mCherry the smallest. In addition, the low temperature slope decreases with photobleaching for all three RFPs. This low temperature linear regime followed by an abrupt onset in the picosecond dynamics is identical to the INS measurements of RMSD (25). Zaccai and coworkers introduced two global flexibility parameters to quantify the RMSD temperature dependence (26): the resilience $k^*$, defined as proportional to the inverse of the slope of the RMSD temperature dependence at low temperatures expressed as $<k^*> = 0.00138/(d<3x^2>/dT)$, where the angular brackets denote an ensemble average over atoms and time; and the dynamical transition temperature $T_D$, where the dynamics sharply change. Here we take a similar approach. Assuming that the THz absorption, which selectively measures the optically active vibrational modes of proteins at picosecond time scale, is proportional to the vibrational density of states (VDOS) denoted as $g(\omega)$ which is in turn derived from the Fourier transform of the atomic velocity autocorrelation: $g(\omega) = <v(0)*v(t)>$ (27), we can use the harmonic approximation to express the absorption coefficient $\alpha(\omega,T)$ and $g(\omega)$ for a particular frequency $\omega$ as:

$$\alpha(\omega,T) \sim g(\omega) = \int_{-\infty}^{\infty} <\sqrt{m}\mathbf{v}(0)\cdot\sqrt{m}\mathbf{v}(t)>e^{i\omega t}dt$$

$$= \int_{-\infty}^{\infty} <\omega\sqrt{m}\mathbf{x}(0)\cdot\omega\sqrt{m}\mathbf{x}(t)>e^{i\omega t}dt$$

$$= m\omega^2 \int_{-\infty}^{\infty} <\mathbf{x}(0)\cdot\mathbf{x}(t)>e^{i\omega t}dt$$

$$\sim m\omega^2<x^2> \sim m\omega^2\frac{k_B T}{k^*}$$

$$\Rightarrow k^* \sim \frac{m\omega^2 k_B}{\partial\alpha(\omega,T)/\partial T} \sim \frac{C(\omega)}{\partial\alpha(\omega,T)/\partial T}$$

where k* is an effective force constant and C(ω) is a frequency dependent coefficient. We refer to $k^*$ as the THz resilience. For a given frequency we evaluate the resilience by the inverse of the slope of the molar absorptivity temperature dependence in the linear regime.

|  | mCherry | TagRFP-T | mOrange2 |
|---|---|---|---|
| $T_{M, Unbleached}$ (K) | 359±2 | 345±1 | 353±4 |
| $T_{M, Bleached}$ (K) | 373±2 | 350±1 | 367±4 |
| $k^*_{Unbleached}$ | 43±14 | 18±1 | 21±3 |
| $k^*_{Bleached}$ | 45±4 | 22±3 | 31±9 |

Table 1. The melting temperature for all samples was determined by the midpoint in the dropoff of far-UV CD signal at 218 nm. The effective force constant $k^*$ is calculated for low temperature regime (100-200K) at 1.0 THz, while other frequencies show similar tendency.

In Table 1 we show the resilience extracted from the low temperature THz molar absorptivity with the measured melting temperatures ($T_M$) from CD measurements. The data show an increase in $T_M$ for the photobleached state of each RFP. Strikingly, the THz resilience follows the thermal stability for all three proteins in both the unbleached and bleached states. This is the first demonstration that THz resilience correlates with thermal stability, and is consistent with previous neutron scattering measurements for hemoglobins from different species (28), suggesting that stronger resilience leading to higher thermal stability may be a general phenomenon.

We now turn to the dynamical transition measurements which show for the first time that the protein turn-on is distinct from the solvent turn-on. The $T_D$ seen in the THz measurements are different for the different proteins, but for all three the $T_D$ increases with photobleaching. There has been no previous report of $T_D$ changing with the functional state of the protein and the result is unexpected. The transition has been ascribed to the slaving of protein dynamics to the solvent (29, 30). At low temperatures, the immobile water prevents large amplitude protein motions. A rapid increase in the solvent mobility at T~200K, enables the protein to access the large amplitude anharmonic motions necessary for physiological function. While the specific activation energies for the solvent motions can vary with the specific protein surface causing $T_D$ variation protein to protein, the $T_D$ change with photobleaching is not expected, as there is little change in the surface solvent exposure. The increase in both the $T_D$ and $T_M$ suggests that possibly the protein structural dynamics are playing a role in the temperature dependence. A simple power law spectral analysis of the THz absorption confirms a significant difference in the photobleached protein dynamics. The results in Figure 3 show that the temperature dependence of the power law substantially changes with photobleaching. The power law factor denotes the index of the power law fit of each THz absorption as a function of frequency. For the unbleached RFP's, the power law is relatively flat below 200 K, and then drops above this temperature, whereas for the photobleached samples the 200 K inflection is nearly absent.

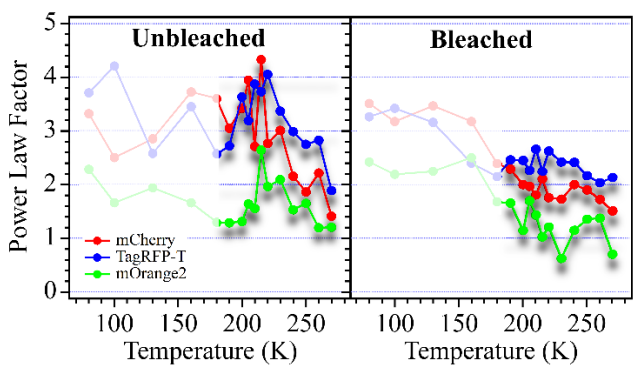

Figure 3. Power law factors from fits to THz absorption frequency dependence for unbleached (left) and bleached (right) RFP samples: mCherry (red), TagRFP-T (blue), and mOrange2 (green) as a function of temperature. The high temperature region above dynamical transition has been emphasized.

To resolve the source of the dynamical transition temperature shift we isolate the protein and solvent contributions to the THz absorbance. In the case of INS measurements, the separation of the solvent dynamics from the protein dynamics is accomplished by successive measurements of protonated protein with deuterated solvent, and deuterated protein with protonated solvent (31, 32). For optical measurements, ideally the isolation of different dynamics can be done using distinct spectral signatures for the protein versus solvent. At room temperature, this is not possible as protein solution THz measurements are dominated by water, which is generally modeled by the sum of three Debye terms with relaxation times of 8.3 ps, 1 ps, and 0.2 ps and an intermolecular water vibration at 5.3 THz (33-35). The relaxational absorption is sufficiently large that intramolecular protein vibrations can be entirely neglected.

At low temperatures, however the THz absorbance changes considerably. The low frequency absorption drops dramatically (36-43) as the water relaxation times increase with decreasing temperature (44) moving the relaxational water absorption loss into the MHz range, while the water intermolecular vibrations (33, 34), centered at 5.3 THz, remains. Thus we can use the 5.3 THz resonance to monitor the temperature dependent solvent dynamics.

To guide our fitting of the protein contribution, we performed molecular dynamics simulations using quasiharmonic mode analysis (QHA) and dipole-dipole autocorrelation calculations. For QHA the solvent is minimized and the harmonic approximation focuses on the collective motions of the hydrated protein only, whereas the dipole-dipole autocorrelation calculations includes all motions contributing to the absorbance (45). The VDOS and autocorrelation results are shown in Figure S2 in the Supplementary Information.

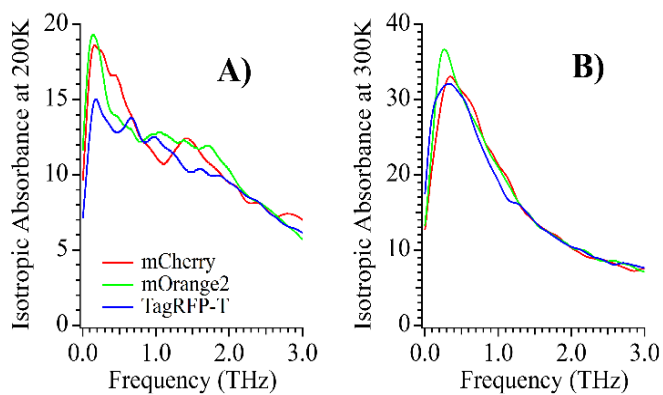

Figure 4. Isotropic THz absorption calculated by quasi-harmonic analysis at 200K (A) and 300K (B). High temperature absorption shows a Lorentzian-like peak centered at ~0.5 THz for all three RFPs. The same color scale is used for both plots.

The QHA calculated isotropic absorptivity for the three RFPs below and above the $T_D$ are shown in Figure 4. There is negligible difference among the RFPs for these calculations. The ordering in the net absorbance seen in Figure 1 is reproduced by the autocorrelation calculations however (see Figure S2 in Supplementary Information). The calculated QHA isotropic absorbance exhibits similar Lorentzian-like absorption peaks centered near 0.5 THz which become narrow and slightly shift up in frequency with increasing temperature.

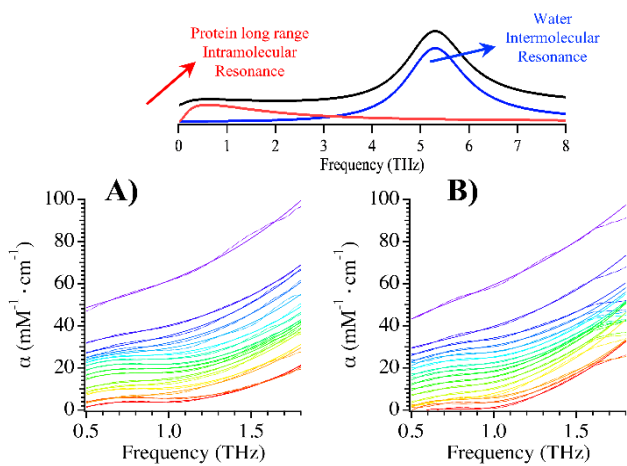

Figure 5. Comparisons of the frequency dependent THz molar absorptivity with resonance fitting lines for photoactive (A) and photobleached (B) TagRFP-T. The offset of 2 mM$^{-1}$cm$^{-1}$ was applied to distinguish different temperatures. The top figure illustrates the decomposing the absorption into two Lorentzian resonances: protein long-range intramolecular resonance centered at ~0.6 THz, and larger-amplitude water intermolecular resonance at 5.3 THz.

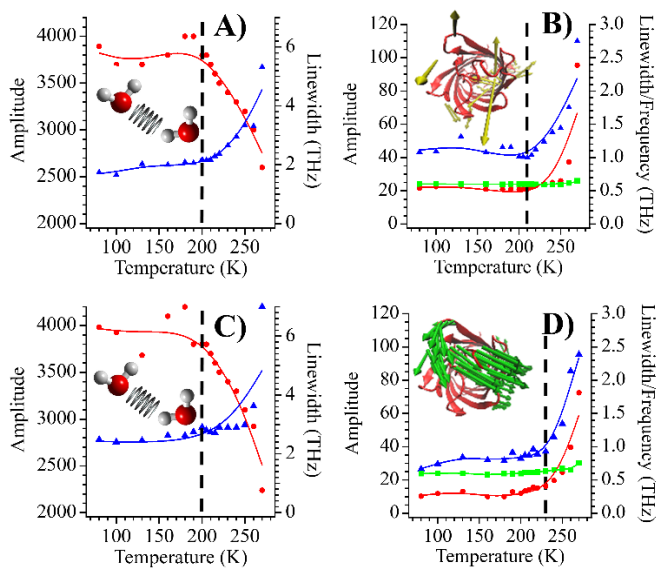

Figure 6. Resonance fitting parameters for photoactive (top) and photobleached (bottom) TagRFP-T. A). C).: Amplitude (circle, red) and linewidth (triangle, blue) corresponding to water intermolecular resonance centered at 5.3 THz. B). D).: Amplitude (circle, red), linewidth (triangle, blue) and central frequency (square, green) of the protein intramolecular resonant band, the photobleached state shows higher collectivity of the motion. The interpolated lines are used as a guide for the eye.

Based on the QHA results we fit our measured absorbance with two Lorentzians with one frequency set at 5.3 THz for the intermolecular water vibration. Note that only the tail of the water resonance contributes to the spectral feature as background absorption and a slight change in central frequency will not alter other fitting parameters. The results for TagRFP-T in Figure 5A and 5B show that the model fits the data well for the unbleached and bleached states. In Figure 6 we see the temperature dependent behavior of the protein resonant band is distinctly different from the water resonance. Specifically, Figure 6A and 6C show the amplitude and linewidth for the 5.3 THz water band for TagRFP-T unbleached and bleached respectively. In both cases the amplitude begins to rapidly decrease and linewidth rapidly begins to increase at 200 K. The solvent transition is unaffected by the bleaching, as one might expect for the solvent dynamics. As the temperature increases the water resonance broadens as the amplitude decreases with the net integrated intensity of the solvent intermolecular excitations remaining constant. The broadening is consistent with additional excitations accessible with increasing thermally activated mobile waters. Figure 6B shows the frequency, amplitude and linewidth of the low frequency band for the unbleached protein. The central frequency is nominally at 0.6 THz and blue shifts at higher temperatures, in agreement with the QHA peak in Figure 4. Both the amplitude and linewidth increase at a transition temperature of 210 K for the unbleached TagRFP-T. That is, the dynamical onset temperature for the low frequency protein motions is higher than that of the solvent. We will define two onset temperatures, from the sharp turn-on points of each curve, $T_{DS}$ for the solvent and $T_{DP}$ for the protein. Figure 6D shows the results for the photobleached TagRFP-T. The photobleached state $T_{DP}$ shifts up to 230 K, while the solvent $T_{DS}$ remains the same. mCherry has the same result (see Figure S4 in Supplementary Information). The picosecond water dynamics turn-on at 200 K for both unbleached and bleached protein, whereas the unbleached $T_{DP}$ is 210 K and bleached is 230 K. Just as the melting temperature shifts up with bleaching, so does the turn-on for the picosecond dynamics. For mOrange2 the solvent transition is again at 200K for both unbleached and bleached protein. For unbleached mOrange2 the protein dynamical turn- on is again at 210 K, but the bleached mOrange2 is substantially different, with the protein dynamical transition nearly absent and a slight inflection at 220K. In all three cases the protein temperature dependence does not follow the solvent.

## IV. DISCUSSION

The correlation between structural resilience and thermal stability is perhaps intuitive and consistent with previous neutron studies, however the correlation of the protein dynamical transition temperature with thermal stability is somewhat surprising. The rapid onset in the protein dynamics with temperature has long been understood to be associated with the thermally activated motions of the surrounding solvent (29), so any trend with protein structural stability is not expected. The protein dynamical transition arises from the need to break hydrogen bonds within the solvent cage to accommodate motions. This phenomenon has been termed the slaving of the protein's dynamics to the solvent. At the same time it is understood that the solvent excitations are influenced by the specific protein surface (30). $T_D$ has been observed to vary for different measurements (25, 46, 47), and for different proteins for a single technique (28). For example, INS measurements using the same energy beamline for different proteins reveal onset temperatures as low as 200K to as high as 250K. This variation has been attributed to differences in the specific protein-solvent surface interaction (32). The direct dependence of the protein dynamical onset on the water mobility onset appeared to be confirmed in two neutron studies where both the solvent transition and protein transition were separately measured (48, 49). For example, while $T_{DS}$ for maltose binding protein (MBP) and hen egg white lysozyme (HEWL) are different, for both cases the $T_{DP}$ coincides with $T_{DS}$. We note that both MBP and HEWL have two lobes surrounding a binding cleft. MBP is almost entirely α-helical with a small β-sheet region at the binding site, whereas HEWL has one lobe that is mainly α-helix and the other mainly β-sheet. The picture that emerged is that for a given protein the $T_{DS}$ is dictated by the average solvent binding energies to the specific protein surface, and the protein structural dynamics follow the thermal activation of these surface-solvent excitations. In the results presented here we see a somewhat extraordinary and different result where the protein dynamical turn-on clearly is different than the solvent turn-on. For each of the three proteins, the solvent $T_{DS}$ remains essentially unchanged in the two photostates, indicating little change in the solvent-protein interactions. This is consistent with structural measurements of KillerRed and IrisFP using similar bleaching conditions (6, 50). For both KillerRed and IrisFP there is little structural change with photobleaching. The slight decrease in THz absorbance also is consistent with the structure remaining intact, as it has been found the low temperature THz absorbance increases substantially with structural loss.

The protein dynamical onsets however are not the same as the solvent onsets and are dependent on photostate.  Even in the unbleached state, $T_{DP}$ is shifted up relative to the solvent $T_{DS}$ for all three proteins. This difference with the previous comparisons of $T_{DP}$ and $T_{DS}$ for MBP and HEWL may in part arise from the more rigid β-barrel structure of the RFPs.  The shift increases dramatically by 20 K with photobleaching for mCherry and TagRFP-T. This shift has not been reported previously and requires a further examination of the solvent slaving idea.  As previously discussed, at low temperatures the water motions are limited, trapping the protein configuration.  As the temperature increases, water cluster motions are thermally activated, lifting the constraints on the protein dynamics.  If in the measurement frequency range the motions are highly localized, then the number of thermally activated mobile water clusters needed for the motions to occur is small, and temperature dependence of the protein dynamics will closely follow the solvent dynamical transition.  Larger populations of mobile water clusters are required to execute delocalized motions, thus leading to an increase in their turn-on temperature relative to $T_{DS}$.  In the picosecond range it has been shown that solvent fluctuations have an Arrhenius temperature dependence (6, 30, 51).  If we simplistically relate the Arrhenius dependence to the thermal population of mobile waters, the protein dynamical onset shift can be used as a measure of required mobile water population needed for the onset of protein motions. Previous RMSD measurements have reported activation energies between 20-40 kJ/mol (51-57). This range of activation energies give a fractional mobile water increase of 100 - 200% ($[N(T_{DP})-N(T_{DS})]/N(T_{DS}) = \Delta N/N$) for the $T_{DS}$ to $T_{DP}$ shift of 200 to 210 K for the unbleached state.  For the photobleached $T_{DP}$ shift to 230 K, the threshold mobile water population increases to 400 - 2200%.  That is in the photobleached state, there is a substantial increase in the required mobile waters to access the THz intramolecular vibrations.  This large increase suggests the bleached state motions are more spatially extended, with more distant regions moving in concert.

The increased collectivity of picosecond-timescale motions in the photobleached state suggested by these THz measurements is consistent with enhanced internal coupling through water channels formed by photoinduced alteration of the internal protein structure.  For each of the two RFP's (6, 50) whose structure has been solved in both unbleached and bleached states, under bleaching conditions similar to those used in our studies, the β-barrel structure is nearly unchanged, however a CAVER analysis shows that additional water channels appear in the photobleached state (see Figure S5 in Supplementary Information).  The dissipation of the excess energy via strong structural fluctuations provides an avenue for the water channel formation.  These additional water channels can provide H-bond coupling within the β-barrel interior.  The impact of the water channels on the collectivity is evident in a comparison between the CAVER water channel maps and the B-factor maps for KillerRed and IrisFP (see Figure S5 in Supplementary Information).  In the photobleached state the B-factor uniformity increases in the same regions as the water channels form.  The enhanced coupling provided by the water channels is also consistent with the increase in thermal stability that we measure.  Finally we note that these same water channels likely are responsible for the loss in fluorescence in the photobleached state.  All organic fluorophores, including RFPs, suffer from irreversible photobleaching after exposure to prolonged and excessive illumination (58).  For RFPs, candidate mechanisms leading to fluorescence loss are oxidation and/or cis-trans isomerization of the chromophore (6, 12, 22).  Under the lower intensity illumination conditions of our study, oxidation is thought to be the dominant mechanism whereas the cis-trans isomerization mechanism occurs for more extreme conditions (59-61).  While oxygen is required for initial chromophore maturation, it has been found that photobleaching for mOrange2, and TagRFP-T is oxygen sensitive, and oxygen-free conditions result in the improved photostability (5).  Photobleaching via oxygen diffusion through the water channel in β7-β10 region in mCherry has also been discussed (22).  The presence of water channels in the photobleached state can explain an increase in dynamical collectivity, an increase in thermal stability and a loss of fluorescence by increasing oxygen access leading to trapping of the chromophore in a protonated state (50, 62-64).

In summary, we find an increase in structural stability and vibrational collectivity with RFP photobleaching, consistent with enhanced intramolecular coupling via internal water channel formation with prolonged photo excitation.  Both the strength of THz absorption and THz low temperature resilience correlate with thermal stability.  The temperature dependent THz absorbance spectra can be used to separate the solvent and protein dynamical onsets.  We find the dynamical onset of the protein motions does not coincide with that of the solvent, and that it increases in the photobleached state. We suggest that the shifting of the protein dynamical onset relative to the solvent arises from the threshold mobile water population needed for the protein motions to be accessible.

## V. ACKNOWLEDGEMENTS

The authors thank Prem P. Chapagain at Florida International University for providing RFP topology and parameter files. This work was made possible by funding from National Science Foundation MRI^2 grant DBI2959989, IDBR grant DBI1556359, MCB grant MCB1616529, the Department of Energy BES grant DE-SC0016317, and the NSF Physics Frontier Center at JILA, NSF PHY 1734006. RJ is a staff member in the Quantum Physics Division of NIST. Certain commercial equipment, instruments, or materials are identified in this paper in order to specify the experimental procedure adequately. Such identification is not intended to imply that the materials or equipment identified are necessarily the best available for the purpose.

## VI. AUTHOR CONTRIBUTIONS

M.X. performed THz TDS, CD, light scattering, fluorescence measurements, calculations, and analyzed data; D.K.G. performed THz TDS, fluorescence measurements, and analyzed data; R.J. provided samples; A.G.M. designed and conceived measurements; and all authors discussed the results and contributed significantly to the writing of the article.

# Supplementary Information

# Photo Switching of Protein Dynamical Collectivity

Mengyang Xu[1], D. K. George[1], R. Jimenez[2], A. G. Markelz[1]
[1]University at Buffalo, SUNY, Buffalo, NY, 14260 USA
[2] University of Colorado, Boulder, CO, 80309 USA

Figure S1 shows the temperature dependent circular dichroism measurements for the unbleached and bleached RFP samples.  $T_m$ is most easily attained by the peak in the first temperature derivative of the ellipticity, however in the case of mCherry, the increase in $T_m$ is beyond the measurement range of the CD instrument.  In this case the lower bound of the Tm is determined by the midpoint between the low temperature value and the value at 373 K.  For all three RFPs, the melting temperature $T_m$ increases with photobleaching.  In both photostates the THz net molar absorptivity and low temperature (<200K) resilience follow the thermal stability of the individual proteins.

Figure S2 shows the results for the dipole autocorrelation calculation of the THz absorbance at room temperature. A time-correlation function of the dipole operator relates the dynamics of an equilibrium ensemble to the absorption coefficient.  This correlation function reflects the spontaneous fluctuations in the dipole moment and contains information on states of system and broadening due to relaxation.  The calculated power spectra by the dipole-dipole autocorrelation function from MD trajectories at room temperature shows the same tendencies as measured: the most flexible TagRFP-T shows the largest calculated THz absorbance, then mOrange2, and mCherry having the smallest calculated absorbance.

Figure S3 shows the improvement in the fitting of the THz absorbance spectra from a simple power law to the sum of two Lorentzians, one centered at 5.3 THz for the known intermolecular water vibration and the second Lorentzian's parameters determined by the fitting.  The second Lorentzian typically has a center frequency of 0.6 THz, which agrees well with our quasiharmonic calculation of the intramolecular vibrations of the protein.

Figure S4 shows the fitting parameters extracted using the double Lorentzian fit as described in the main text for mCherry.  The solvent dynamical transition $T_{DS}$ is independent of photo state for both mCherry and mOrange2, as was seen for TagRFP-T in the main text.  For the photobleached mCherry, the dynamical turn on temperature for the protein increases by ~20K, as was seen for TagRFP-T.  The temperature dependence of the low frequency absorbance band of the photobleached state for mOrange2 also substantially changes however rather than just a net shift in $T_{DP}$, the transition is almost absent, and the temperature dependence appears linear.

Figure S5 shows how changes in the uniformity of the relative B factors correlate with the presence of water channels present in the bleached state of killer Red.  Fig. S5 A) and B) show the change in the structure and relative B factor in the photoactive and bleached states respectively.  Figs. S5 C) and D) show the calculated water channels based on these structural measurements. The CAVER software package (1) is applied to calculate the water channels. The calculations find that the water channels substantially increase in the photobleached state as was reported previously in an X-ray crystallographic structure study (2). At the same time we note here that in the structural regions where these channels form, the B-factor uniformity increases, suggesting a stabilization and increase in collectivity in these regions.

Figure S6 shows the fluorescence peak of mOrange as a function of concentration. The peak shifted from ~565 nm to ~610 nm with concentration change of two orders.

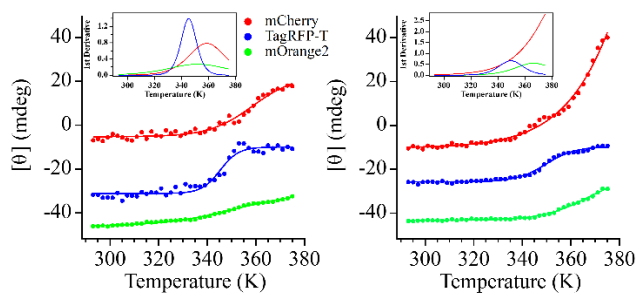

Figure S1. Melting measurements of unbleached A) and bleached B) for mCherry (red), TagRFP-T (blue), and mOrange2 (green) using far-UV CD spectra with sigmoidal fitting ranging from 20°C to 100°C. Insets show the first derivative of the temperature dependence. The peak of the derivative indicates the melting temperature Tm. The values are consistent with the reported $T_m$ = 356K of enhanced green fluorescent protein (EGFP), which has a photostability similar to mOrange2.

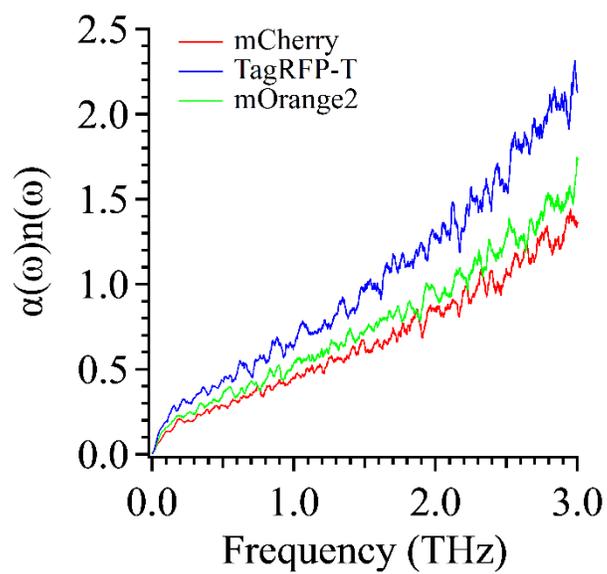

Figure S2. The frequency dependent $\alpha(\omega)n(\omega)$ for mCherry (red), TagRFP-T (blue), and mOrange2 (green) at the same hydration level (h($g_{water}/g_{protein}$) ~ 1.8), reproduced by dipole-dipole autocorrelation analysis at room termperature within the same THz frequency range of the experiments. The tendency resembles the order of thermal stability.

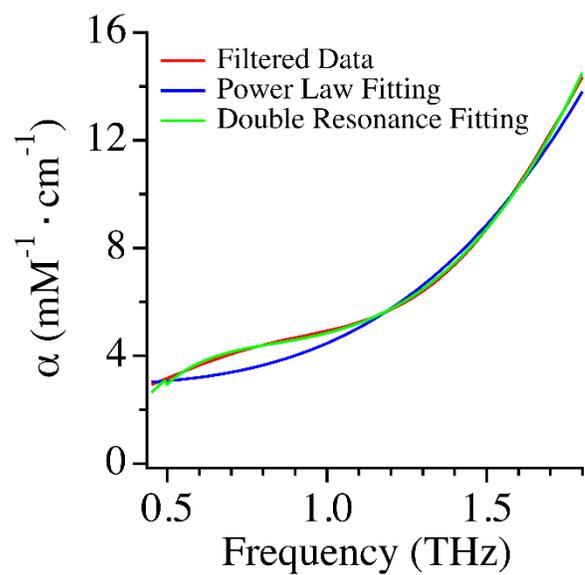

Figure S3. THz molar absorptivity (red) with power-law (blue), and double resonance (green) fittings for unbleached mCherry at 80K.

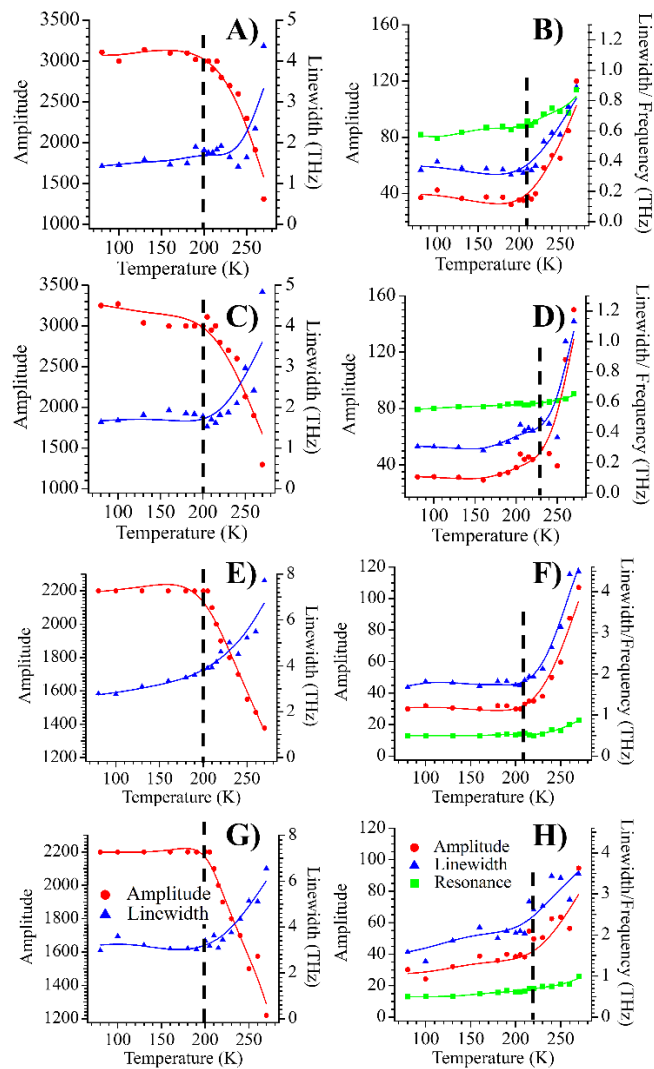

Figure S4. Temperature dependent parameters from double-resonance fitting to terahertz absorbance spectra. mCherry photoactive and photobleached parameters are shown in (A,B) and (C,D) respectively. mOrange2 photoactive and photobleached parameters are shown in (E,F) and (G,H) respectively. The amplitude (red circles) and linewidth (blue trianges) of the 5.3 THz water resonance is shown in the left column (A, C, E and G). The amplitude (red circles), linewidth (blue triangles) and resonant frequency (green squares) for the protein resonant band are shown in the right column (B,D, F and H). The interpolated lines are used as a guide to the eye.

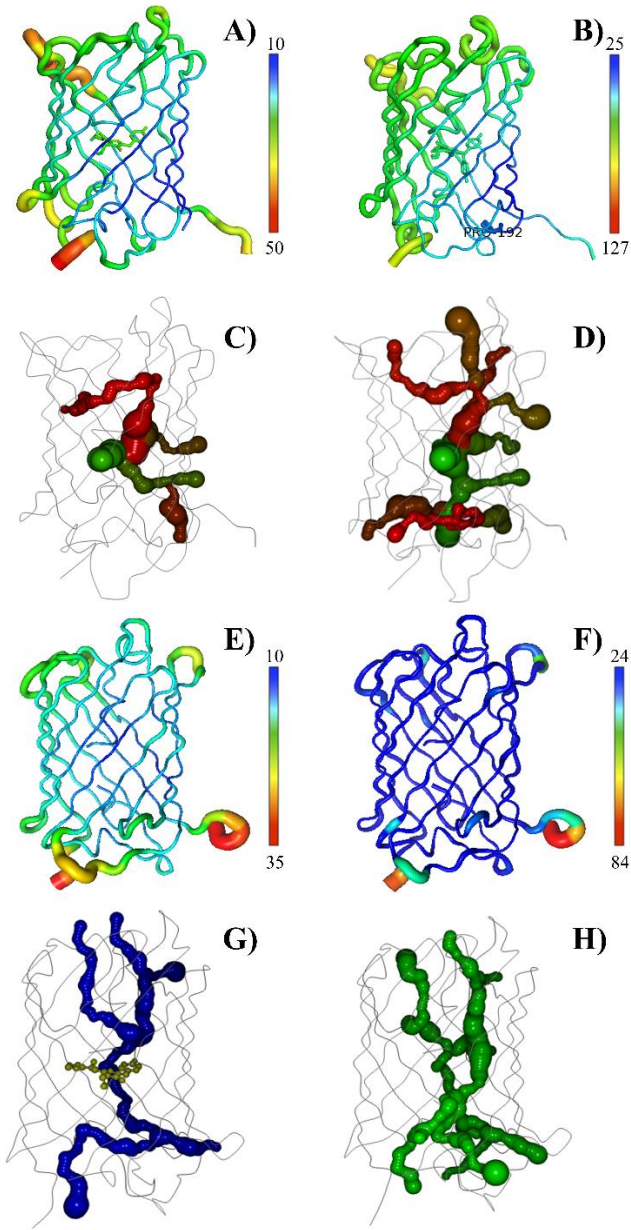

Figure S5. Debye-Waller B factor surface plots for native A) and photobleached B) KillerRED using 2WIQ.pdb and 2WIS.pdb; and for native E) and photobleached F) IrisFP using 2VVH.pdb and 4LJD.pdb. The width and color of the ribbon reflect the average B-value dependent on residues. Note that the colors have been normalized to their average value to focus on relative B-factor distribution. The observed additional water channel gated by Residue 192 is marked in (B). C), D), G) and H) show the water channels calculated by CAVER, corresponding to native and photobleached states for KillerRED and IrisFP, respectively. In both cases, bleached states show additional water channels as well as more uniform B-factor variance.

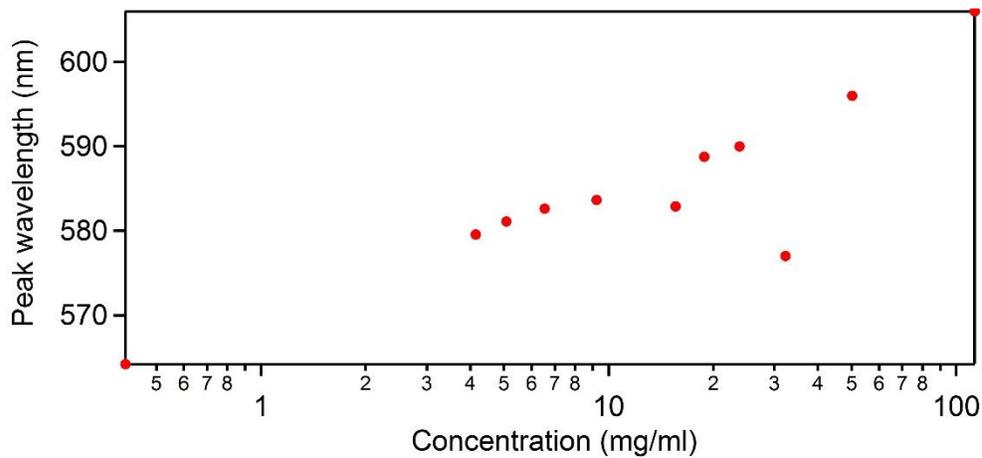

Figure S6. Fluorescence peak of mOrange at room temperature as a function of concentration